\documentclass[twocolumn,
    showpacs,
    amsmath,
    amssymb,
    nofootinbib,
    nobibnotes,
    letterpaper,
    citeautoscript,
    10pt,
    pra]{revtex4-1}
\usepackage{amsfonts}
\usepackage[english]{babel}
\usepackage[T1]{fontenc}
\usepackage{times}
\usepackage{mathrsfs}
\usepackage{graphicx}
\usepackage{dcolumn}
\usepackage{bm}
\usepackage[colorlinks,bookmarks=true,citecolor=blue,linkcolor=red,urlcolor=blue]{hyperref}
\usepackage[tight, FIGTOPCAP, hang, raggedright, nooneline]{subfigure}

\usepackage{color}

\newcommand{\bb}[1]{\mathbf{#1}}
\newcommand{\del}[0]{\partial}
\renewcommand{\d}[1]{\textrm{d}#1}

\begin{document}

\title{Quantum transport in Dirac materials: Signatures of tilted and anisotropic Dirac and Weyl cones}
\author{Maximilian Trescher, Bj\"orn Sbierski, Piet W. Brouwer and  Emil J. Bergholtz}
\affiliation{Dahlem Center for Complex Quantum Systems and Institut f\"ur Theoretische Physik, Freie Universit\"at Berlin, Arnimallee 14, 14195 Berlin, Germany}
\date{\today}

\begin{abstract}
We calculate conductance and noise for quantum transport at the nodal point for arbitrarily tilted and anisotropic Dirac or Weyl cones. Tilted and anisotropic dispersions are generic in the absence of certain discrete symmetries, such as particle-hole and lattice point group symmetries. Whereas anisotropy affects the conductance $g$, but leaves the Fano factor $F$ (the ratio of shot noise power and current) unchanged, a tilt affects both $g$ and $F$. 
Since $F$ is a universal number in many other situations, this finding is remarkable.
We apply our general considerations to specific lattice models of strained graphene and a pyrochlore Weyl semimetal.
\end{abstract}

\maketitle

\section{Introduction} 
Driven by a combination of key advances in materials fabrication and profound conceptual progress, the past decade has witnessed an explosive increase in the study of electronic systems dispersing linearly around isolated band touching points \cite{wehling_dirac_2014}.
Notably, this includes graphene \cite{novoselov_electric_2004,novoselov_two-dimensional_2005}, various two-dimensional organic compounds \cite{goerbig_tilted_2008}, and the surface states of three-dimensional topological insulators \cite{fu_topological_2007,moore_topological_2007,roy_topological_2009,zhang_crossover_2010}.
In three dimensions, a Dirac semimetal \cite{fang_anomalous_2003,murakami_phase_2007}, which has two coinciding linear band touching points with opposite chirality, was observed experimentally \cite{liu_discovery_2014,liu_stable_2014}, and the first Weyl semimetal \cite{wan_topological_2011,hosur_recent_2013}, which has non-degenerate band-touching points, was observed very recently \cite{xu_experimental_2015,lu_experimental_2015,lv_discovery_2015}.
Subsequently the first transport measurements on Weyl semimetals were performed \cite{huang_observation_2015,zhang_observation_2015}.
A Weyl semimetal phase is also predicted to occur, e.g., in multilayer structures \cite{burkov_weyl_2011} and pyrochlore iridates \cite{wan_topological_2011}.

By virtue of stochiometry, the Fermi level lies exactly at the nodal point of the low-energy ``cones'' in many of these materials, and their electronic behavior is neither that of insulators --- there is no gap --- nor that of conventional metals --- there is a vanishing density of states at the nodal point. Indeed it has been shown experimentally \cite{novoselov_two-dimensional_2005,zhang_experimental_2005} and theoretically \cite{tworzydlo_sub-poissonian_2006,katsnelson_zitterbewegung_2006,baireuther_quantum_2014} that the conductivity $\sigma$ reaches a minimal but finite value at a nodal point in two dimensions, whereas a nodal point in three dimensions is characterized by a finite conductance $G$, its conductivity $\sigma$ being zero \cite{baireuther_quantum_2014,sbierski_quantum_2014}. The Fano factor $F$, defined as the ratio of shot noise power and current, was found to be an excellent indicator of the quantum nature of electronic transport at the nodal point, taking the universal sub-Poissonian value $F=1/3$ in graphene \cite{tworzydlo_sub-poissonian_2006}. In Weyl semimetals $F$ was found to discriminate between a pseudoballistic regime \cite{baireuther_quantum_2014} at weak disorder and a diffusive regime at strong disorder \cite{sbierski_quantum_2014}. Unlike the conductance $G$, which retains a dependence on the ratio $W/L$ of sample width $W$ and sample length $L$, the Fano factor $F$ is independent of both $W$ and $L$.

Anisotropy and tilt of the cones are often neglected, essentially for two distinct reasons: (i) they are forbidden by symmetry in important special cases, such as graphene, and (ii) they do not alter the topology of the low-energy theory. 
Here, however, we demonstrate that tilts and, to a lesser extent, anisotropies lead to clear signatures in quantum transport, affecting both the conductance and the Fano factor in absence of disorder. We find the tilt dependence of the Fano factor $F$ remarkable, because in many cases of interest $F$ was found to be a number with a considerable degree of universality \cite{beenakker_suppression_1992,jalabert_universal_1994,baranger_mesoscopic_1994,tworzydlo_sub-poissonian_2006,baireuther_quantum_2014}.
Our results apply --- with various degrees of numerical relevance --- to a number of experimentally relevant systems for which tilted and anisotropic conical dispersion either occur generically, 
as in the case of Weyl semimetals,
or for which the forbidding symmetries are easily broken, such as strained graphene.

\section{Tilted and anisotropic cones}
\begin{figure}[t]
    \begin{center}
        \includegraphics[width=0.35\textwidth]{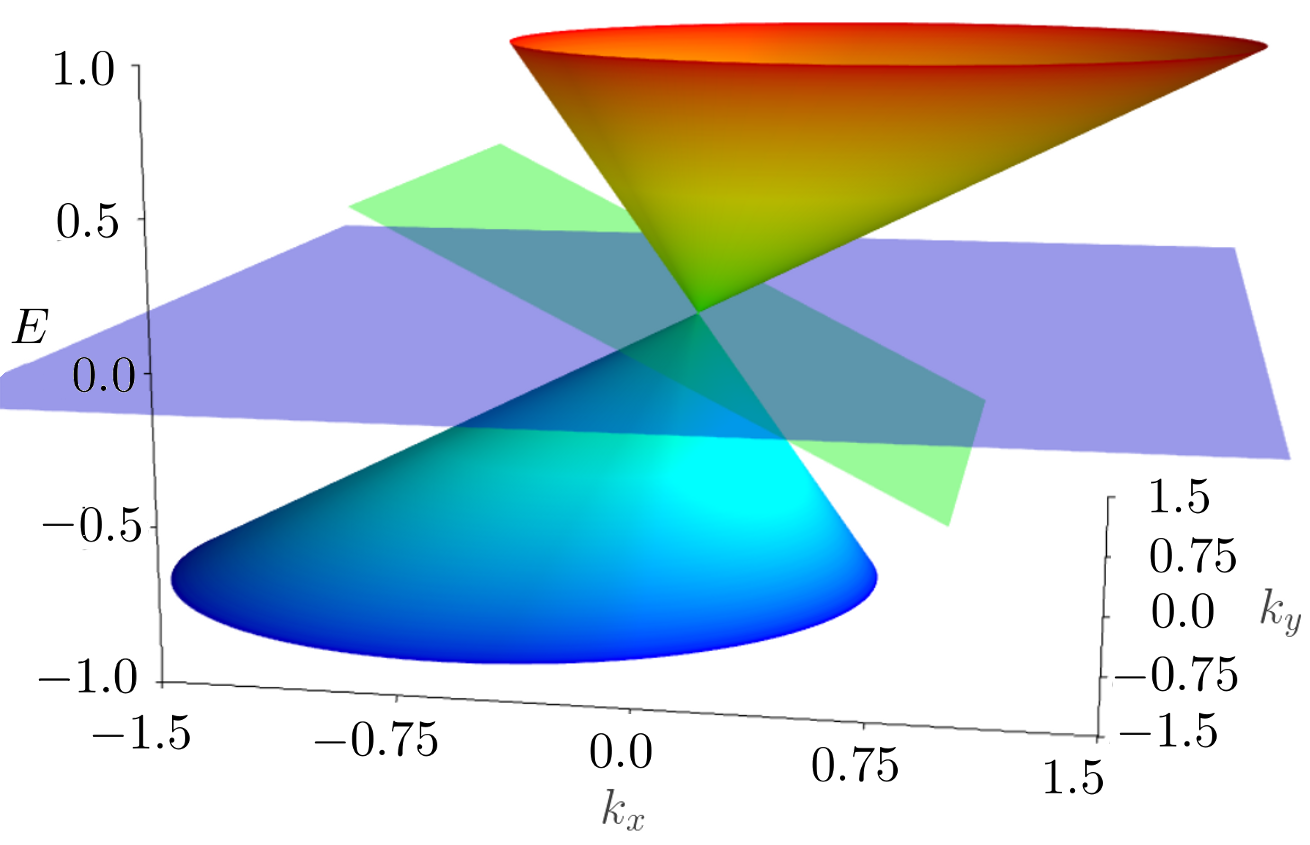}
    \end{center}
    \caption{(Color online) A tilted Dirac cone in two dimensions. The momentum coordinates are labeled $k_x$ and $k_y$; the third dimension represents energy.
    Transparent planes indicate zero-energy plane (violet) and tilt ${\mathbf a} = (-0.5, 0)$ (green plane), respectively.}
    \label{fig:tilted-dirac-visual}
\end{figure}

In the vicinity of a nodal point, a generic Dirac or Weyl Hamiltonian can be written as
\begin{align}
     H &= \sum_{i,j} v_{ij} k_i \bb{\sigma}_j + (a_i k_i - u ) \sigma_0, \label{lowEH}
\end{align}
where the sum is over $i,j = x,y$ or $i,j = x,y,z$ for dimensionalities $d=2$ and $d=3$, respectively. Further $\sigma_{x,y,z}$ are the Pauli matrices and $\sigma_0$ is the $2 \times 2$ unit matrix. The dispersion is shown schematically in Fig.\ \ref{fig:tilted-dirac-visual}. The ``tilt'' term proportional to $a_i$ is typically discarded, as it does not affect the eigenspinors and, hence, the topology of the band structure. As we show below, inclusion of this term does affect transport at the nodal point, however.
Tilts can occur only if particle-hole symmetry is absent, and tilt is additionally constrained by point group symmetries. With a suitable choice of the pseudospin quantization axis, the anisotropy matrix $v_{ij}$ can be brought to upper diagonal form, $v_{yx} = v_{zx} = v_{zy} = 0$. Anisotropies are generic if the cone is not located at a high symmetry point in the Brillouin zone.

Considering graphene as an important example, we note that the trigonal ``warping'' of Dirac cones respects the crystalline symmetries and leads to anisotropies, but only at quadratic order in the momentum $k$. The anisotropies to linear order (\ref{lowEH}), which amount to a ``squeezing'' of the cone along some direction, are, just as any tilt of the cone, forbidden by the threefold point group symmetry of the honeycomb lattice, combined with the location of the Dirac cones at high-symmetry points in the Brillouin zone. However, as soon as the threefold rotation symmetry is relaxed anisotropies occur. If, in addition, second-nearest-neighbor hopping is also taken into account the particle-hole symmetry is lost, and the cones acquire finite tilts. This scenario applies to strained graphene and will be discussed in more detail below. For three-dimensional Weyl semimetals, the band touching occur at lower symmetry points; hence anisotropies and tilts are ubiquitous.  

\section{Transport: low-energy theory}

We calculate the conductance $G$ and the Fano factor $F$ for a region of length $L$ and width $W$, taking the limit $W \gg L$ in order to eliminate a spurious dependence on the transverse boundary conditions \cite{tworzydlo_sub-poissonian_2006}. We choose the $x$ axis in the transport direction, so that the nodal semimetal corresponds to the region $0 < x < L$, whereas the source and drain leads have $x < 0$ and $x > L$, respectively. The potential $u$ is set to zero for $0 < x < L$, to model transport at the nodal point. We take the limit $u \to \infty$ for $x < 0$ and $x > L$ to model strongly doped leads.

The transverse momentum $k_{\perp} = k_y$ ($d=2$) or $k_{\perp} = (k_y,k_z)$ ($d=3$) is conserved, and for each value of $k_{\perp}$ we calculate the transmission coefficient $T(k_{\perp})$ by matching wave functions in the sample and the leads (see Appendix for details). The conductance $G$ per cone is then given by the Landauer formula
\begin{align}
    G =& \frac{e^2}{h} \left( \frac{W}{2 \pi} \right)^{d-1} \int  \d^{d-1}{k_{\perp}} T(k_{\perp}).
    \label{the-int}
\end{align}
The aspect-ratio dependence can be partially eliminated by changing to the dimensionless conductance referred to a cube, defined by the relations
\begin{equation}
  G = \frac{e^2}{h} \left( \frac{W}{L} \right)^{d-1} g.
    \label{conductance-g}
\end{equation}
In two dimensions $g$ is identical to the conductivity $\sigma$. 
In three dimensions, a finite value for $g$ in the limit $W$, $L \to \infty$ implies a vanishing conductivity $\sigma = G L/W^2 = (e^2/h) g/L$.
The Fano factor, the ratio of shot noise and current, is given by \cite{buettiker_scattering_1990}
\begin{align}
    F = \dfrac{\int \d^{d-1} k_{\perp} T(k_{\perp})(1-T(k_{\perp}))}{
  \int  \d^{d-1} k_{\perp} T(k_{\perp}) }.
    \label{fano}
\end{align}

\begin{figure}[b]
    \begin{center}
        \includegraphics[width=0.5\textwidth]{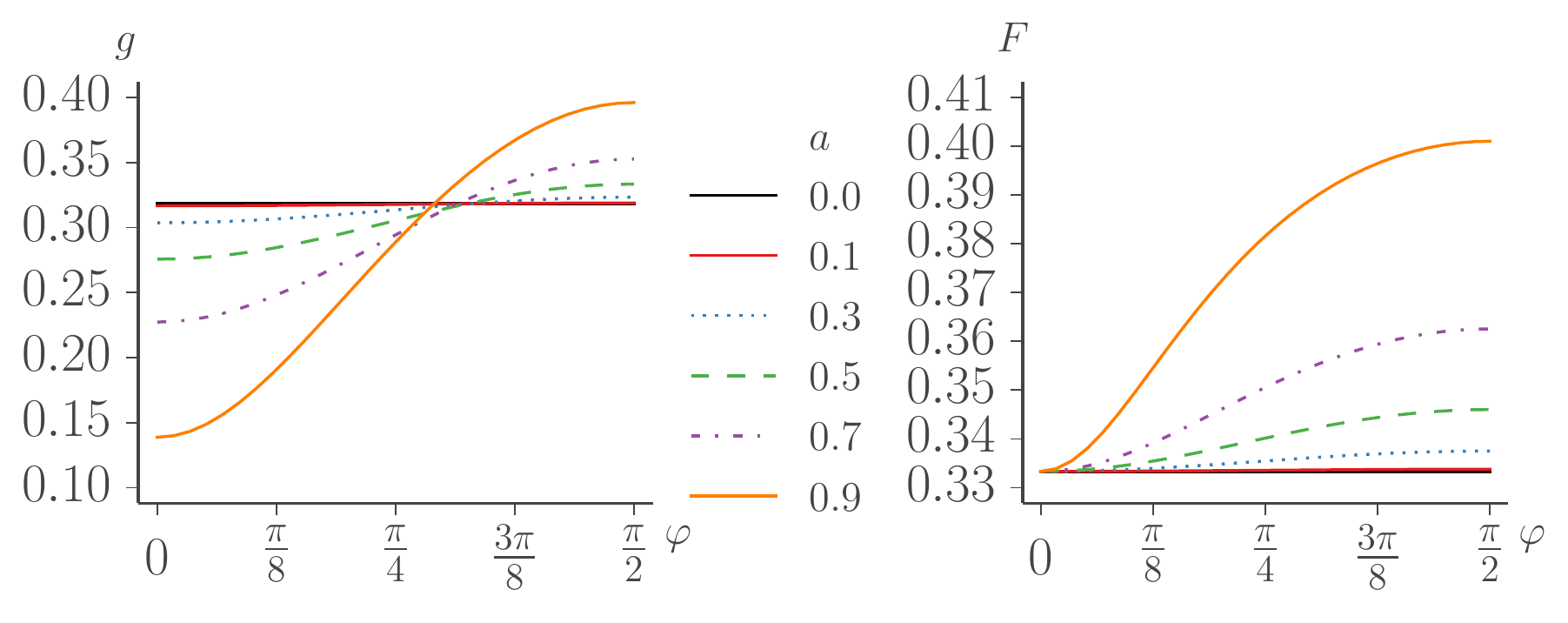}
    \end{center}
    \caption{(Color online) Dimensionless conductance $g$ and Fano factor $F$ for a tilted two-dimensional Dirac cone, as a function of the angle $\varphi$ between transport and tilt direction. The tilt strength $a$ is given in the legend.}
    \label{fig:dirac-tilt}
\end{figure}

{\it No anisotropy, no tilt.--}
For the isotropic cone without tilt ($a_i = 0, v_{ij} = v_0 \delta_{ij}$) the conductance and Fano factor are known from the literature \cite{tworzydlo_sub-poissonian_2006,sbierski_quantum_2014,baireuther_quantum_2014},
\begin{align}
  g &= \frac{1}{\pi}, \quad\quad F = \frac{1}{3}, & (d=2), 
    \label{isotropic-results-dirac} \\
  g &= \frac{\ln 2}{2 \pi},\quad F = \frac{1 + 2 \ln 2}{6 \ln 2}, & (d=3).
  \label{isotropic-results-weyl}
\end{align}

{\it Anisotropy, no tilt.--} For the general anisotropic case but without tilt, $a_{i} = 0$, $i=1,\ldots,d$, one finds 
\begin{align}
  g &= \frac{1}{\pi} \frac{v_{xx}^2 + v_{xy}^2}{v_{xx} v_{yy}}
  & (d=2),
\end{align}
while the Fano factor is unaffected by the anisotropy; i.e., $F$ is given by Eq.\ (\ref{isotropic-results-dirac}). For the diagonal case ($v_{xy} = 0$) this result can be understood as a simple scaling of the $y$ coordinate, which affects the conductance $g$, but not the Fano factor $F$.
In three dimensions the exact result in the diagonal case ($v_{xy} = v_{xz} = v_{yz} = 0$) is given by the corresponding rescaling
\begin{align}
    g &= \frac{\ln 2}{2 \pi} \frac{v_{xx}^2}{v_{yy}v_{zz}} & (d=3),
    \label{}
\end{align}
while there is no simple formula for the general case. Still, the Fano factor remains unaffected by any anisotropy and is given by Eq. (\ref{isotropic-results-weyl}).

\begin{figure}[t]
    \includegraphics[width=0.48\textwidth]{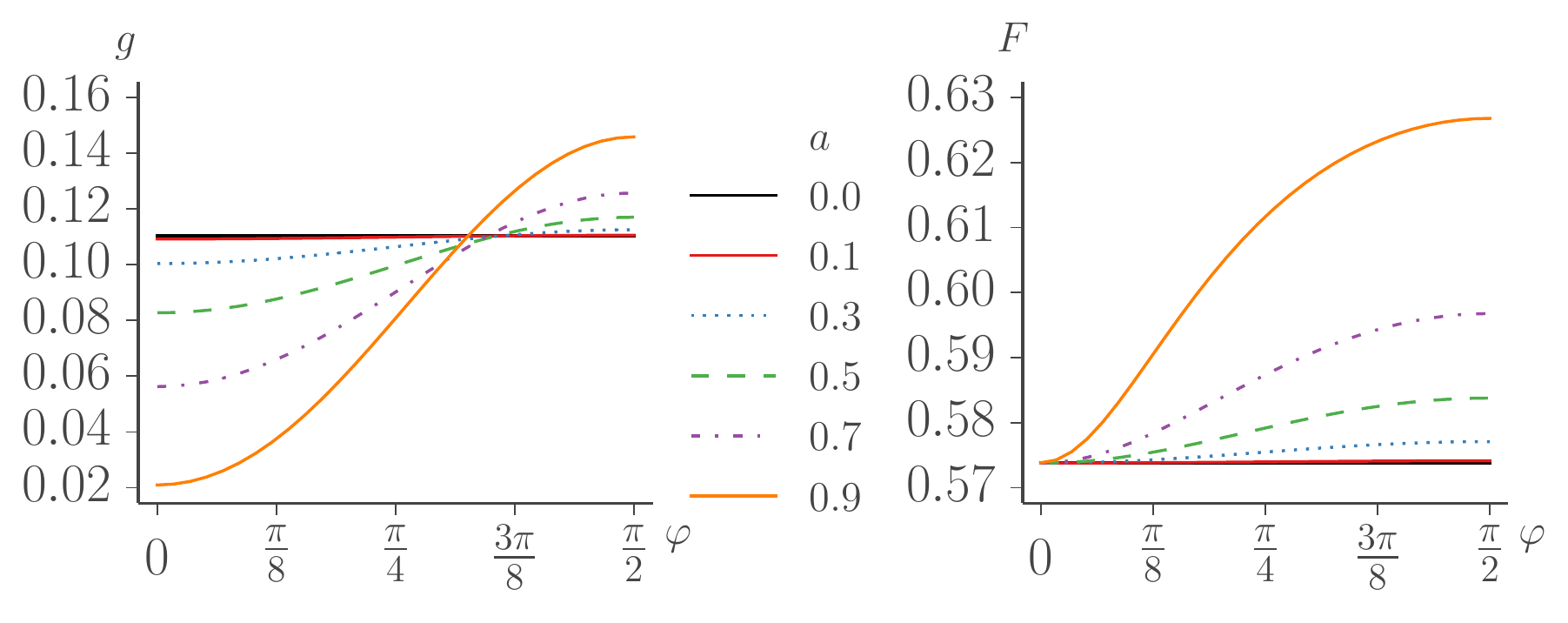}
    \caption{(Color online) Same as in Fig.\ \ref{fig:dirac-tilt}, but for a three-dimensional Weyl cone.}
    \label{fig:weyl-tilted}
\end{figure}

{\it No anisotropy, tilted cones.--} Although a closed analytical solution for a tilted Dirac cone is possible in two dimensions, the explicit expressions are too lengthy to be reproduced here. Instead, we will present a numerical evaluation of the solution for representative values of the tilt parameters $a_x$, $a_y$, and $a_z$ for fixed values of $v_{ij} = \delta_{ij}$. Without anisotropy the dimensionless conductance $g$ and the Fano factor depend on the total magnitude $a^2 = a_x^2 + a_y^2$ ($d=2$) or $a^2 = a_x^2 + a_y^2 + a_z^2$ of the tilt and the angle $\varphi = \arccos(|a_x|/a)$ between the tilt axis and the transport direction only. The limit $a = 1$ corresponds to a maximally tilted cone with a flat band along the tilt direction. Results are shown in Figs.\ \ref{fig:dirac-tilt} and \ref{fig:weyl-tilted} for $d=2$ and $d=3$ and for representative values of the tilt strength $a$. 

We note that the results are quantitatively different in two and three dimensions, but qualitatively very similar. There is an important difference between a tilt parallel to the transport direction, where $g$ decreases upon increasing the tilt strength, and a tilt perpendicular to the transport direction, where $g$ increases with increasing tilt strength. The Fano factor $F$ is unaffected by a tilt in the transport direction, and increases with increasing tilt if there is a finite angle between the tilt direction and the transport axis. Interestingly, upon averaging over all orientations of the tilt axis, we find a systematic but small decrease in conductance for both two and three dimensions. The main conclusion, however, is that the Fano factor is no longer a universal number once the tilt of the dispersion is taken into account, but depends on the magnitude and direction of the tilt. 

The analytical solution for a two-dimensional Dirac cone takes a simple form if the tilt axis and the transport direction are collinear ($\varphi = 0$). In that case one finds $g = (1/\pi) \sqrt{1 - a^2}$, $F = 1/3$. Further, for small tilt strengths it is possible to expand the analytical solution in two dimensions. We find
\begin{align}
    g &= \frac{1}{\pi} +
    \frac{a^2}{2 \pi} \left( \frac{4}{3} \sin^2 \varphi - 1 \right) + \mathcal{O}(a^4), \\
    F &= \frac{1}{3} + \frac{2 a^2}{45} \sin^2 \varphi + \mathcal{O}(a^4),
    \label{ay-results}
\end{align}
which deviates less then $1\%$ from the exact value up to $a = 0.5$. 

{\it Anisotropy and tilted cones.--} In the presence of both anisotropy and tilt the dimensionless conductance and the Fano factor are qualitatively similar as in the absence of anisotropy.
However, for a tilt in transport direction the Fano factor changes if the anisotropies are not orientated along the axis of the reference frames, i.e., if one of $v_{xy}, v_{xz}$ or $v_{yz}$ is nonzero.

\section{Application to lattice models} In generic lattice models, the cones are both anisotropic and tilted, and moreover, contributions from an even number of cones must be taken into account simultaneously.  Below, we provide explicit results for two specific tight-binding models. 

{\it Strained graphene.--} In ``intrinsic,'' unstrained graphene the Dirac cones are located at high symmetry points in the Brillouin zone.
The application of strain changes the positions of the Dirac points and the cones are no longer protected by crystalline symmetries. Whereas the simplest tight-binding model with nearest-neighbor hopping only is particle-hole symmetric, which rules out a tilt of the Dirac cones, realistic tight-binding models have longer-range hopping, which lifts the particle-hole symmetry \cite{charlier_tight-binding_1991}. As an example, we now apply the above calculations to the model of quinoid-type strained graphene, as described by Goerbig {\em et al.} in Ref. \onlinecite{goerbig_tilted_2008}. Transport properties of strained graphene have been studied earlier \cite{pellegrino_transport_2011}, but without the inclusion of a tilt of the Dirac cones. 

\begin{figure}[t]
    \includegraphics[width=0.4\textwidth]{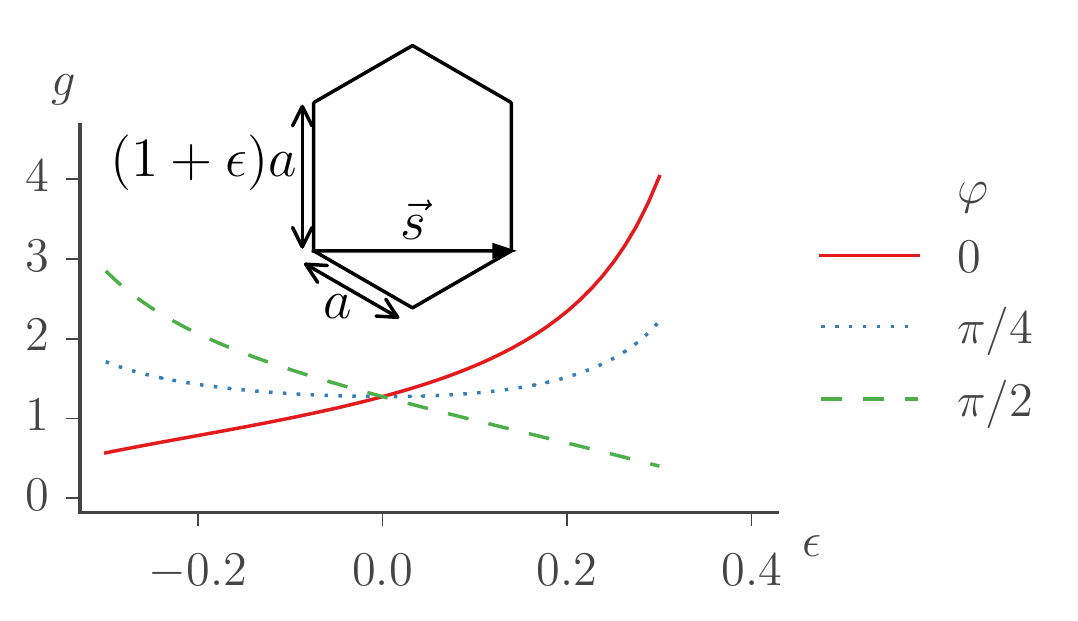}
    \caption{(Color online) Dimensionless conductance $g$ of a strained graphene sheet as a function of the strain $\epsilon$ for different orientations $\varphi$ (the angle between transport direction and $\vec{s}$) and summed over both Dirac cones and spin. The inset shows a hexagon of the graphene lattice for $\epsilon = 0.2$. }
    \label{fig:graphene}
\end{figure}

A schematic of the tight-binding model for quinoid-type strained graphene is shown in the inset of Fig.\ \ref{fig:graphene}. It consists of a honeycomb lattice which is extended/compressed in the direction perpendicular to the lattice vector $\vec{s}$, such that each hexagon has four ``short'' bonds of length $a$ and two ``long'' bonds of length $a'$ for positive strain $\epsilon > 0$. Strain is measured in terms of a dimensionless strain parameter $\epsilon = a'/a-1$. The tight-binding model of Ref.\ \onlinecite{goerbig_tilted_2008} contains nearest-neighbor hopping amplitudes as well as next-nearest-neighbor hopping, and we take the magnitudes of the hopping amplitudes from Ref.\ \onlinecite{goerbig_tilted_2008}.
Figure \ref{fig:graphene} shows the conductance $g$ for strains $0 < |\epsilon| < 0.3$ and three representative angles $\varphi$. The strain is perpendicular to the $\vec{s}$ direction (as depicted in Fig. \ref{fig:graphene}). The angle $\varphi$ is defined as the angle between the transport direction and $\vec{s}$.
The tilt is of order $a/v \sim 0.06$ ($v$ being the velocity in tilt direction) for the (already quite unrealistic) strain $\epsilon = 0.3$ \cite{goerbig_tilted_2008}. As a consequence of this numerically small value of the tilt strength, the relative change in Fano factor remains small, $\lesssim 0.1\%$ for $\epsilon < 0.3$. While this variation is probably out of reach of experimental detection, it shows that Fano factor $F=1/3$ is not strictly ``universal'' in graphene but can be changed by the breaking of symmetries.

\begin{figure}[t]
    \begin{minipage}[c]{0.23\textwidth}
        \includegraphics[width=\textwidth]{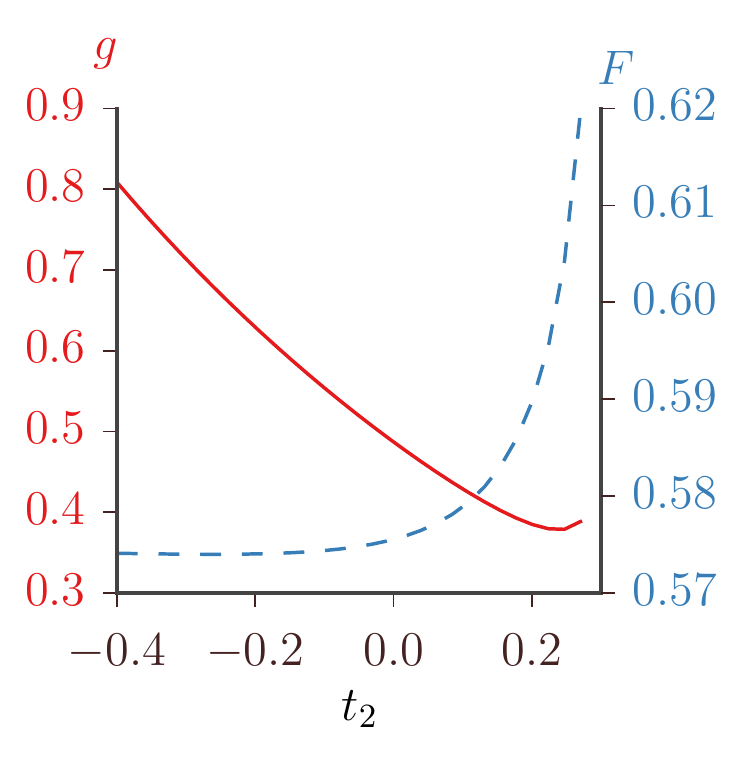}
    \end{minipage}
    \begin{minipage}[c]{0.23\textwidth}
        \centering
        \includegraphics[width=0.8\textwidth]{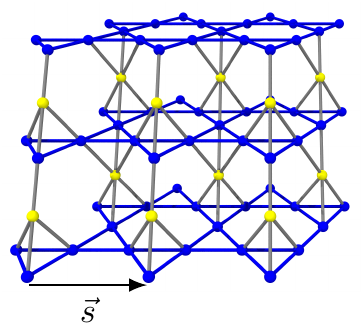}\\
        \includegraphics[width=0.7\textwidth]{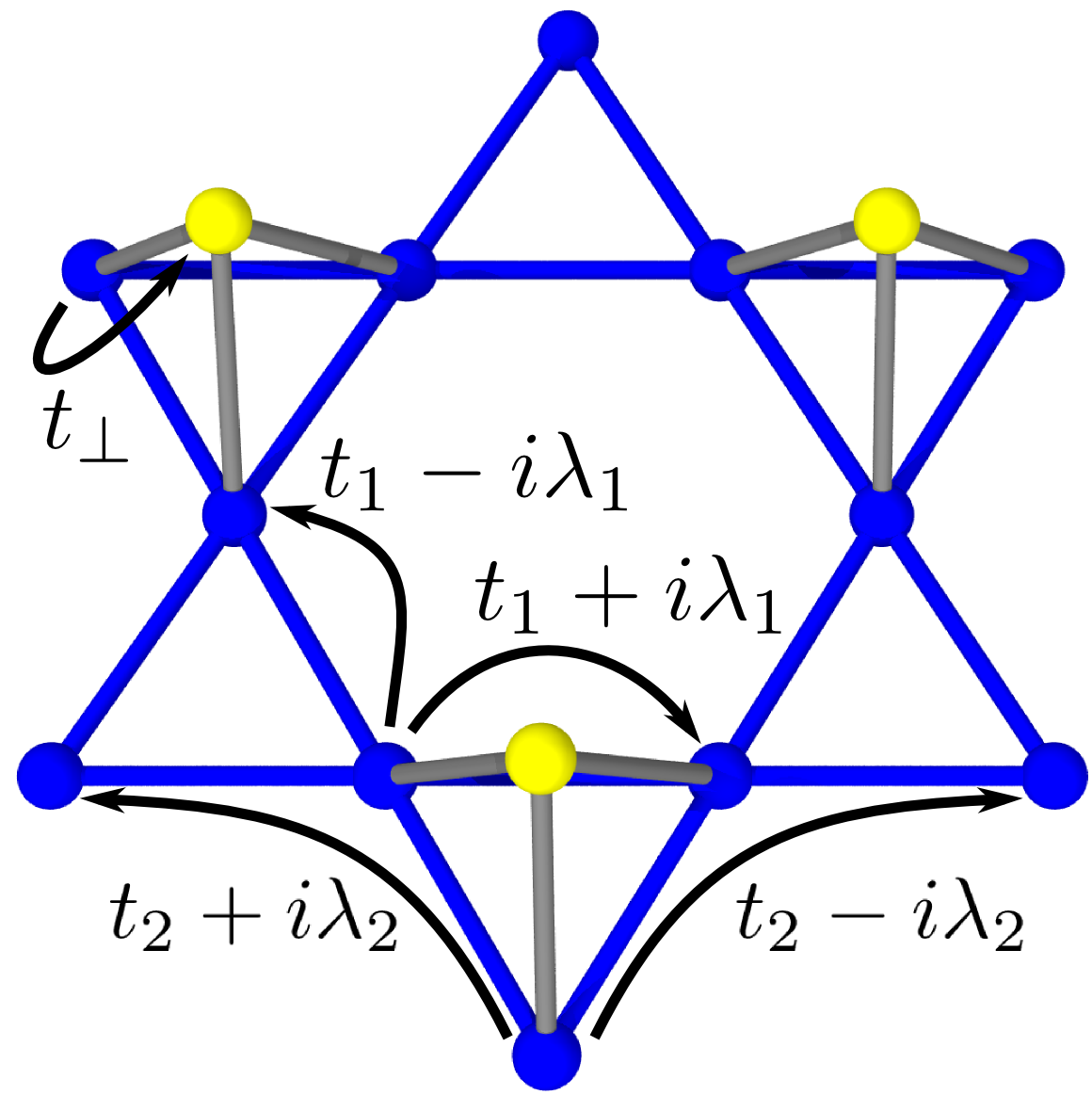}
    \end{minipage}
    \caption{(Color online) Dimensionless conductance $g$ (solid line) and Fano factor $F$ (dashed line) for a pyrochlore slab as a function of the in-plane next-nearest-neighbor hopping amplitude $t_2$ (left panel). Transport direction is parallel to the crystal axis $\vec{s}$ (as shown in the upper right panel). Hopping parameters are indicated in the lower right panel.}
    \label{fig:pyrochlore}
\end{figure}

{\it Weyl semimetal.--} As an example in three dimensions we consider a tight-binding model of a spin-orbit coupled pyrochlore slab which hosts a Weyl semimetal phase with Weyl cones that may be significantly tilted \cite{trescher_flat_2012,bergholtz_topology_2015}. In this case the lattice structure is layered, see Fig.\ \ref{fig:pyrochlore}, so that it is intrinsically anisotropic and no external strain needs to be applied in order to lift any symmetries forbidding a tilt of the Weyl cone. The model consists of a tight-binding Hamiltonian that contains spin-orbit coupling, in-plane and interplane nearest-neighbor hopping amplitudes, and in-plane next-nearest-neighbor hopping amplitudes. It was found to have a Weyl-semimetal phase for a certain range of parameter space, with a tilt of the Weyl cone that depends on the magnitude of the next-nearest-neighbor hopping amplitude $t_2$. There are six Weyl cones, located on the $\Gamma$--M lines \cite{bergholtz_topology_2015} in the projected two-dimensional Brilluoin zone of the slab geometry. The Weyl points are related to each other by the sixfold symmetry of the underlying lattice. We have numerically determined the position as well as tilt and anisotropy parameters as a function of the next-nearest-neighbor hopping $t_2$, keeping the other model parameters, defined in the lower right panel of Fig. \ref{fig:pyrochlore}, fixed ($t_1=-1, t_\perp=2, \lambda_1 = 0.3, \lambda_2=0.2$), and calculated the dimensionless conductance $g$ and the Fano factor $F$. The results are shown in Fig.\ \ref{fig:pyrochlore} for an in-plane transport direction aligned with one of the crystal axes as indicated in the inset. The dependence on the orientation of the pyrochlore slab is very weak, less than $1\%$ for both $g$ and $F$, which can be understood as a consequence of there being six different contributing cones: when rotating the sample, some cones are rotated ``away'' from the transport direction, while others are rotated ``towards'' the transport direction. The changes in transport properties in different cones then have opposite signs (cf. Fig. \ref{fig:weyl-tilted}), leading to a very weak angular dependence of $g$ and $F$. The magnitude of the dimensionless conductance $g$ and the Fano factor $F$ can however differ substantially from the values calculated in the absence of a tilt.

\section{Discussion} 
We have investigated the effect of anisotropies and tilts of Dirac and Weyl cones on quantum transport properties at the nodal point. Neither anisotropies nor tilts change the topology of the band structure and for this reason they are often neglected. We showed that a tilt nevertheless affects the dimensionless conductance $g$ and Fano factor $F$. The latter observation is remarkable, since the Fano factor is often found to be a universal number, that does not depend on system-specific details.

Applying our results to the example of strained graphene, we found that the inclusion of a tilt of the Dirac cone leads to a sizable directional dependence of the conductance $g$. For realistic strains, the tilt effect on the Fano factor $F$ is nonzero, though numerically very small --- underlining the symmetry protected nature of ``universal'' quantum transport in two-dimensional Dirac materials. The consequences of a tilted dispersion, including a shift of the Fano factor $F$, may be more significant for other two-dimensional materials possessing more strongly tilted Dirac cones, such as the organic compound $\alpha$--(BEDT-TTF)$_2$I$_3$ \cite{goerbig_tilted_2008,katayama_pressure-induced_2006}.

While the first observations of Weyl semimetals are a great experimental success, they all observed time-reversal symmetric Weyl semimetals \cite{xu_observation_2015,lu_experimental_2015,lv_discovery_2015,huang_observation_2015,zhang_observation_2015}. These Weyl semimetals have additional states crossing the Fermi level opposed to the still hypothetical time-reversal symmetry breaking Weyl semimetals \cite{wan_topological_2011},
where the only states crossing the Fermi level are Weyl nodes, as is the case in our example of a pyrochlore slab.
Hence the full transport properties could be obtained from the combined contribution of the Weyl nodes, whereas in the experimentally observed materials one should account also for the other states at the Fermi level.

In a recent work three of us
proposed that the Fano factor can be used as a universal quantity to discriminate different transport regimes in a disordered Weyl semimetal \cite{sbierski_quantum_2014}: In a calculation that did not include tilt or anisotropy, $F$ was found to take the ballistic value (\ref{isotropic-results-weyl}) below a critical disorder strength, whereas $F$ approaches the {\em smaller} diffusive value $F=1/3$ at larger disorder. Our present results indicate that there is no universal value for the Fano factor in the ballistic limit. However, we also find that a tilt of the Weyl cone can only increase $F$, so that the Fano factor continues to be a powerful indicator discriminating the pseudoballistic and diffusive regimes.

For tilted and anisotropic cones the conductance varies strongly with transport direction
and can be either higher or lower than the conductance of the symmetric cone.
In contrast the Fano factor is only sensitive to the tilt of the cone and, whereas it still depends on the angle between tilt and transport direction, the Fano factor always increases for tilted cones. These insights should be useful for the experimental identification and characterization of a range of Weyl and Dirac materials by means of transport measurements.

\acknowledgments
We acknowledge related discussions with Teemu Ojanen and Masafumi Udagawa. E.J.B. and M.T. are supported by
DFG\textquoteright{}s Emmy Noether program (BE 5233/1-1). This work was in part supported
by the Helmholtz VI ``New States of Matter and Their Excitations.''

\appendix
\section{}
We calculate ballistic transport in a scattering region of length $L$ and width $W$ as described in the main text.
The Hamiltonian is given by
\begin{align}
    H &= \sum_{ij} v_{ij} k_i \bb{\sigma}_j + (a_i k_i - u ) \sigma_0 \,.
\end{align}
The $a_i k_i$ terms can be interpreted as a tilt of the cone and $v_{ij}$ as a $d\times d$ matrix describing the anisotropy of the dispersion, where it is sufficient to have nonzero entries on the upper triangular to describe all possible anisotropies of a cone.
The dispersion is given by
\begin{widetext}
\begin{align}
    \epsilon_{1,2}(k) &= -u + a_{x} k_{x} + a_{y} k_{y} + a_{z} k_{z} \nonumber\\
    &\pm \sqrt{k_{x}^{2} ( v_{xx}^{2} + v_{xy}^{2} + v_{xz}^{2})
    + 2 k_{x} (k_{y} (v_{xy} v_{yy} + v_{xz} v_{yz}) + k_{z} v_{xz} v_{zz})
    + k_{y}^{2} (v_{yy}^{2} + v_{yz}^{2})
    + 2 k_{y} k_{z} v_{yz} v_{zz} + k_{z}^{2} v_{zz}^{2}} \,.
    \label{dispersion}
\end{align}
\end{widetext}
Whenever the above square-root expression occurs we will abbreviate it as $\sqrt{\cdots}$.
Then the spinors are

\begin{align}
    \chi_{1,2} &= \left(\begin{matrix}- \dfrac{k_{x} v_{xx} - i \left(k_{x} v_{xy} + k_{y} v_{yy}\right)}{k_{x} v_{xz} + k_{y} v_{yz} + k_{z} v_{zz} \mp \sqrt{\cdots}}\\1\end{matrix}\right)
    \label{spinors}
\end{align}
and the velocities used to normalize incoming and outgoing plane waves are
$v(k) = \del_k \epsilon(k)$.
We consider the limit of highly doped leads ($u \rightarrow \infty$) in $u = 0$ in the scattering region.
The transverse momentum $k_{\perp} = k_y$ ($d=2$) and $k_{\perp} = (k_y, k_z)$ ($d=3$) is quantized due to the finite width $W$,
\begin{align}
    k_y &= \frac{2 \pi n}{W} &
    k_z &= \frac{2 \pi m}{W} \,.
    \label{qs}
\end{align}
For any  mode of given momentum $k_y, k_z$ we determine the $x$ component of the wave vectors $k_{\mathrm{in}}, k_r, k_t$ of the incoming, reflected, and transmitted wave
and the $x$ component in the scattering region $\tilde{k}_{1,2}$ by solving Eq. \eqref{dispersion} for $\epsilon(k) = 0$.
Then we calculate the transmission and reflection amplitude by wave function matching at the beginning and the end of the scattering region ($x=0, L$):
\begin{align}
    \frac{1}{\sqrt{v_{\mathrm{in}}}} \chi_{\mathrm{in}} +  \frac{r}{\sqrt{v_r}} \chi_r &= \alpha \chi_1 + \beta \chi_2 \,,\\
    \frac{t}{\sqrt{v_{t}}} \chi_{t} &= \alpha \chi_1 \exp^{i\tilde{k}_1 L} + \beta \chi_2 \exp^{i\tilde{k}_2 L}\,.
    \label{wavefunction-matching}
\end{align}
The total transmission probability can be obtained by summing over all modes:
\begin{align}
    T &= \sum_{k_{\perp}} |t(k_{\perp})|^2 \,.
    \label{transmission-proba}
\end{align}
In the limit $\frac{L}{W}\rightarrow 0$ one may replace the sum by an integral which gives Eq. (2) of the main text.

\begin{thebibliography}{30}%
\makeatletter
\providecommand \@ifxundefined [1]{%
 \@ifx{#1\undefined}
}%
\providecommand \@ifnum [1]{%
 \ifnum #1\expandafter \@firstoftwo
 \else \expandafter \@secondoftwo
 \fi
}%
\providecommand \@ifx [1]{%
 \ifx #1\expandafter \@firstoftwo
 \else \expandafter \@secondoftwo
 \fi
}%
\providecommand \natexlab [1]{#1}%
\providecommand \enquote  [1]{``#1''}%
\providecommand \bibnamefont  [1]{#1}%
\providecommand \bibfnamefont [1]{#1}%
\providecommand \citenamefont [1]{#1}%
\providecommand \href@noop [0]{\@secondoftwo}%
\providecommand \href [0]{\begingroup \@sanitize@url \@href}%
\providecommand \@href[1]{\@@startlink{#1}\@@href}%
\providecommand \@@href[1]{\endgroup#1\@@endlink}%
\providecommand \@sanitize@url [0]{\catcode `\\12\catcode `\$12\catcode
  `\&12\catcode `\#12\catcode `\^12\catcode `\_12\catcode `\%12\relax}%
\providecommand \@@startlink[1]{}%
\providecommand \@@endlink[0]{}%
\providecommand \url  [0]{\begingroup\@sanitize@url \@url }%
\providecommand \@url [1]{\endgroup\@href {#1}{\urlprefix }}%
\providecommand \urlprefix  [0]{URL }%
\providecommand \Eprint [0]{\href }%
\providecommand \doibase [0]{http://dx.doi.org/}%
\providecommand \selectlanguage [0]{\@gobble}%
\providecommand \bibinfo  [0]{\@secondoftwo}%
\providecommand \bibfield  [0]{\@secondoftwo}%
\providecommand \translation [1]{[#1]}%
\providecommand \BibitemOpen [0]{}%
\providecommand \bibitemStop [0]{}%
\providecommand \bibitemNoStop [0]{.\EOS\space}%
\providecommand \EOS [0]{\spacefactor3000\relax}%
\providecommand \BibitemShut  [1]{\csname bibitem#1\endcsname}%
\let\auto@bib@innerbib\@empty
\makeatother
\bibitem[{\citenamefont {Wehling}\ \emph {et~al.}(2014)\citenamefont
  {Wehling}, \citenamefont {Black-Schaffer},\ and\ \citenamefont
  {Balatsky}}]{wehling_dirac_2014}%
  \BibitemOpen
  \bibfield  {author} {\bibinfo {author} {\bibfnamefont {T.}~\bibnamefont
  {Wehling}}, \bibinfo {author} {\bibfnamefont {A.}~\bibnamefont
  {Black-Schaffer}}, \ and\ \bibinfo {author} {\bibfnamefont {A.}~\bibnamefont
  {Balatsky}},\ }\href {http://dx.doi.org/10.1080/00018732.2014.927109}
  {\bibfield  {journal} {\bibinfo  {journal} {Adv. Phys.}\ }\textbf {\bibinfo
  {volume} {63}},\ \bibinfo {pages} {1} (\bibinfo {year} {2014})}\BibitemShut
  {NoStop}%
\bibitem[{\citenamefont {Novoselov}\ \emph {et~al.}(2004)\citenamefont
  {Novoselov}, \citenamefont {Geim}, \citenamefont {Morozov}, \citenamefont
  {Jiang}, \citenamefont {Zhang}, \citenamefont {Dubonos}, \citenamefont
  {Grigorieva},\ and\ \citenamefont {Firsov}}]{novoselov_electric_2004}%
  \BibitemOpen
  \bibfield  {author} {\bibinfo {author} {\bibfnamefont {K.~S.}\ \bibnamefont
  {Novoselov}}, \bibinfo {author} {\bibfnamefont {A.~K.}\ \bibnamefont {Geim}},
  \bibinfo {author} {\bibfnamefont {S.~V.}\ \bibnamefont {Morozov}}, \bibinfo
  {author} {\bibfnamefont {D.}~\bibnamefont {Jiang}}, \bibinfo {author}
  {\bibfnamefont {Y.}~\bibnamefont {Zhang}}, \bibinfo {author} {\bibfnamefont
  {S.~V.}\ \bibnamefont {Dubonos}}, \bibinfo {author} {\bibfnamefont {I.~V.}\
  \bibnamefont {Grigorieva}}, \ and\ \bibinfo {author} {\bibfnamefont {A.~A.}\
  \bibnamefont {Firsov}},\ }\href
  {http://www.sciencemag.org/content/306/5696/666} {\bibfield  {journal}
  {\bibinfo  {journal} {Science}\ }\textbf {\bibinfo {volume} {306}},\ \bibinfo
  {pages} {666} (\bibinfo {year} {2004})}\BibitemShut {NoStop}%
\bibitem[{\citenamefont {Novoselov}\ \emph {et~al.}(2005)\citenamefont
  {Novoselov}, \citenamefont {Geim}, \citenamefont {Morozov}, \citenamefont
  {Jiang}, \citenamefont {Katsnelson}, \citenamefont {Grigorieva},
  \citenamefont {Dubonos},\ and\ \citenamefont
  {Firsov}}]{novoselov_two-dimensional_2005}%
  \BibitemOpen
  \bibfield  {author} {\bibinfo {author} {\bibfnamefont {K.~S.}\ \bibnamefont
  {Novoselov}}, \bibinfo {author} {\bibfnamefont {A.~K.}\ \bibnamefont {Geim}},
  \bibinfo {author} {\bibfnamefont {S.~V.}\ \bibnamefont {Morozov}}, \bibinfo
  {author} {\bibfnamefont {D.}~\bibnamefont {Jiang}}, \bibinfo {author}
  {\bibfnamefont {M.~I.}\ \bibnamefont {Katsnelson}}, \bibinfo {author}
  {\bibfnamefont {I.~V.}\ \bibnamefont {Grigorieva}}, \bibinfo {author}
  {\bibfnamefont {S.~V.}\ \bibnamefont {Dubonos}}, \ and\ \bibinfo {author}
  {\bibfnamefont {A.~A.}\ \bibnamefont {Firsov}},\ }\href
  {http://www.nature.com/nature/journal/v438/n7065/abs/nature04233.html}
  {\bibfield  {journal} {\bibinfo  {journal} {Nature}\ }\textbf {\bibinfo
  {volume} {438}},\ \bibinfo {pages} {197} (\bibinfo {year}
  {2005})}\BibitemShut {NoStop}%
\bibitem [{\citenamefont {Goerbig}\ \emph {et~al.}(2008)\citenamefont
  {Goerbig}, \citenamefont {Fuchs}, \citenamefont {Montambaux},\ and\
  \citenamefont {Piechon}}]{goerbig_tilted_2008}%
  \BibitemOpen
  \bibfield  {author} {\bibinfo {author} {\bibfnamefont {M.~O.}\ \bibnamefont
  {Goerbig}}, \bibinfo {author} {\bibfnamefont {J.-N.}\ \bibnamefont {Fuchs}},
  \bibinfo {author} {\bibfnamefont {G.}~\bibnamefont {Montambaux}}, \ and\
  \bibinfo {author} {\bibfnamefont {F.}~\bibnamefont {Piechon}},\ }\href
  {http://link.aps.org/doi/10.1103/PhysRevB.78.045415} {\bibfield  {journal}
  {\bibinfo  {journal} {Phys. Rev. B}\ }\textbf {\bibinfo {volume} {78}},\
  \bibinfo {pages} {045415} (\bibinfo {year} {2008})}\BibitemShut {NoStop}%
\bibitem [{\citenamefont {Fu}\ \emph {et~al.}(2007)\citenamefont {Fu},
  \citenamefont {Kane},\ and\ \citenamefont {Mele}}]{fu_topological_2007}%
  \BibitemOpen
  \bibfield  {author} {\bibinfo {author} {\bibfnamefont {L.}~\bibnamefont
  {Fu}}, \bibinfo {author} {\bibfnamefont {C.~L.}\ \bibnamefont {Kane}}, \ and\
  \bibinfo {author} {\bibfnamefont {E.~J.}\ \bibnamefont {Mele}},\ }\href
  {http://link.aps.org/doi/10.1103/PhysRevLett.98.106803} {\bibfield  {journal}
  {\bibinfo  {journal} {Phys. Rev. Lett.}\ }\textbf {\bibinfo {volume} {98}},\
  \bibinfo {pages} {106803} (\bibinfo {year} {2007})}\BibitemShut {NoStop}%
\bibitem [{\citenamefont {Moore}\ and\ \citenamefont
  {Balents}(2007)}]{moore_topological_2007}%
  \BibitemOpen
  \bibfield  {author} {\bibinfo {author} {\bibfnamefont {J.~E.}~\bibnamefont
  {Moore}}\ and\ \bibinfo {author} {\bibfnamefont {L.}~\bibnamefont
  {Balents}},\ }\href {http://link.aps.org/doi/10.1103/PhysRevB.75.121306}
  {\bibfield  {journal} {\bibinfo  {journal} {Phys. Rev. B}\ }\textbf {\bibinfo
  {volume} {75}},\ \bibinfo {pages} {121306} (\bibinfo {year}
  {2007})}\BibitemShut {NoStop}%
\bibitem [{\citenamefont {Roy}(2009)}]{roy_topological_2009}%
  \BibitemOpen
  \bibfield  {author} {\bibinfo {author} {\bibfnamefont {R.}~\bibnamefont
  {Roy}},\ }\href {http://link.aps.org/doi/10.1103/PhysRevB.79.195322}
  {\bibfield  {journal} {\bibinfo  {journal} {Phys. Rev. B}\ }\textbf {\bibinfo
  {volume} {79}},\ \bibinfo {pages} {195322} (\bibinfo {year}
  {2009})}\BibitemShut {NoStop}%
\bibitem [{\citenamefont {Zhang}\ \emph {et~al.}(2010)\citenamefont {Zhang},
  \citenamefont {He}, \citenamefont {Chang}, \citenamefont {Song},
  \citenamefont {Wang}, \citenamefont {Chen}, \citenamefont {Jia},
  \citenamefont {Fang}, \citenamefont {Dai}, \citenamefont {Shan},
  \citenamefont {Shen}, \citenamefont {Niu}, \citenamefont {Qi}, \citenamefont
  {Zhang}, \citenamefont {Ma},\ and\ \citenamefont
  {Xue}}]{zhang_crossover_2010}%
  \BibitemOpen
  \bibfield  {author} {\bibinfo {author} {\bibfnamefont {Y.}~\bibnamefont
  {Zhang}}, \bibinfo {author} {\bibfnamefont {K.}~\bibnamefont {He}}, \bibinfo
  {author} {\bibfnamefont {C.-Z.}\ \bibnamefont {Chang}}, \bibinfo {author}
  {\bibfnamefont {C.-L.}\ \bibnamefont {Song}}, \bibinfo {author}
  {\bibfnamefont {L.-L.}\ \bibnamefont {Wang}}, \bibinfo {author}
  {\bibfnamefont {X.}~\bibnamefont {Chen}}, \bibinfo {author} {\bibfnamefont
  {J.-F.}\ \bibnamefont {Jia}}, \bibinfo {author} {\bibfnamefont
  {Z.}~\bibnamefont {Fang}}, \bibinfo {author} {\bibfnamefont {X.}~\bibnamefont
  {Dai}}, \bibinfo {author} {\bibfnamefont {W.-Y.}\ \bibnamefont {Shan}},
  \bibinfo {author} {\bibfnamefont {S.-Q.}\ \bibnamefont {Shen}}, \bibinfo
  {author} {\bibfnamefont {Q.}~\bibnamefont {Niu}}, \bibinfo {author}
  {\bibfnamefont {X.-L.}\ \bibnamefont {Qi}}, \bibinfo {author} {\bibfnamefont
  {S.-C.}\ \bibnamefont {Zhang}}, \bibinfo {author} {\bibfnamefont {X.-C.}\
  \bibnamefont {Ma}}, \ and\ \bibinfo {author} {\bibfnamefont {Q.-K.}\
  \bibnamefont {Xue}},\ }\href
  {http://www.nature.com/nphys/journal/v6/n8/abs/nphys1689.html} {\bibfield
  {journal} {\bibinfo  {journal} {Nat. Phys.}\ }\textbf {\bibinfo {volume}
  {6}},\ \bibinfo {pages} {584} (\bibinfo {year} {2010})}\BibitemShut {NoStop}%
\bibitem [{\citenamefont {Fang}\ \emph {et~al.}(2003)\citenamefont {Fang},
  \citenamefont {Nagaosa}, \citenamefont {Takahashi}, \citenamefont {Asamitsu},
  \citenamefont {Mathieu}, \citenamefont {Ogasawara}, \citenamefont {Yamada},
  \citenamefont {Kawasaki}, \citenamefont {Tokura},\ and\ \citenamefont
  {Terakura}}]{fang_anomalous_2003}%
  \BibitemOpen
  \bibfield  {author} {\bibinfo {author} {\bibfnamefont {Z.}~\bibnamefont
  {Fang}}, \bibinfo {author} {\bibfnamefont {N.}~\bibnamefont {Nagaosa}},
  \bibinfo {author} {\bibfnamefont {K.~S.}\ \bibnamefont {Takahashi}}, \bibinfo
  {author} {\bibfnamefont {A.}~\bibnamefont {Asamitsu}}, \bibinfo {author}
  {\bibfnamefont {R.}~\bibnamefont {Mathieu}}, \bibinfo {author} {\bibfnamefont
  {T.}~\bibnamefont {Ogasawara}}, \bibinfo {author} {\bibfnamefont
  {H.}~\bibnamefont {Yamada}}, \bibinfo {author} {\bibfnamefont
  {M.}~\bibnamefont {Kawasaki}}, \bibinfo {author} {\bibfnamefont
  {Y.}~\bibnamefont {Tokura}}, \ and\ \bibinfo {author} {\bibfnamefont
  {K.}~\bibnamefont {Terakura}},\ }\href
  {http://www.sciencemag.org/content/302/5642/92} {\bibfield  {journal}
  {\bibinfo  {journal} {Science}\ }\textbf {\bibinfo {volume} {302}},\ \bibinfo
  {pages} {92} (\bibinfo {year} {2003})}\BibitemShut {NoStop}%
\bibitem [{\citenamefont {Murakami}(2007)}]{murakami_phase_2007}%
  \BibitemOpen
  \bibfield  {author} {\bibinfo {author} {\bibfnamefont {S.}~\bibnamefont
  {Murakami}},\ }\href {http://iopscience.iop.org/1367-2630/9/9/356} {\bibfield
   {journal} {\bibinfo  {journal} {New J. Phys.}\ }\textbf {\bibinfo {volume}
  {9}},\ \bibinfo {pages} {356} (\bibinfo {year} {2007})}\BibitemShut {NoStop}%
\bibitem [{\citenamefont {Liu}\ \emph {et~al.}(2014{\natexlab{a}})\citenamefont
  {Liu}, \citenamefont {Zhou}, \citenamefont {Zhang}, \citenamefont {Wang},
  \citenamefont {Weng}, \citenamefont {Prabhakaran}, \citenamefont {Mo},
  \citenamefont {Shen}, \citenamefont {Fang}, \citenamefont {Dai},
  \citenamefont {Hussain},\ and\ \citenamefont {Chen}}]{liu_discovery_2014}%
  \BibitemOpen
  \bibfield  {author} {\bibinfo {author} {\bibfnamefont {Z.~K.}\ \bibnamefont
  {Liu}}, \bibinfo {author} {\bibfnamefont {B.}~\bibnamefont {Zhou}}, \bibinfo
  {author} {\bibfnamefont {Y.}~\bibnamefont {Zhang}}, \bibinfo {author}
  {\bibfnamefont {Z.~J.}\ \bibnamefont {Wang}}, \bibinfo {author}
  {\bibfnamefont {H.~M.}\ \bibnamefont {Weng}}, \bibinfo {author}
  {\bibfnamefont {D.}~\bibnamefont {Prabhakaran}}, \bibinfo {author}
  {\bibfnamefont {S.-K.}\ \bibnamefont {Mo}}, \bibinfo {author} {\bibfnamefont
  {Z.~X.}\ \bibnamefont {Shen}}, \bibinfo {author} {\bibfnamefont
  {Z.}~\bibnamefont {Fang}}, \bibinfo {author} {\bibfnamefont {X.}~\bibnamefont
  {Dai}}, \bibinfo {author} {\bibfnamefont {Z.}~\bibnamefont {Hussain}}, \ and\
  \bibinfo {author} {\bibfnamefont {Y.~L.}\ \bibnamefont {Chen}},\ }\href
  {http://www.sciencemag.org/content/343/6173/864} {\bibfield  {journal}
  {\bibinfo  {journal} {Science}\ }\textbf {\bibinfo {volume} {343}},\ \bibinfo
  {pages} {864} (\bibinfo {year} {2014}{\natexlab{a}})}\BibitemShut {NoStop}%
\bibitem [{\citenamefont {Liu}\ \emph {et~al.}(2014{\natexlab{b}})\citenamefont
  {Liu}, \citenamefont {Jiang}, \citenamefont {Zhou}, \citenamefont {Wang},
  \citenamefont {Zhang}, \citenamefont {Weng}, \citenamefont {Prabhakaran},
  \citenamefont {Mo}, \citenamefont {Peng}, \citenamefont {Dudin},
  \citenamefont {Kim}, \citenamefont {Hoesch}, \citenamefont {Fang},
  \citenamefont {Dai}, \citenamefont {Shen}, \citenamefont {Feng},
  \citenamefont {Hussain},\ and\ \citenamefont {Chen}}]{liu_stable_2014}%
  \BibitemOpen
  \bibfield  {author} {\bibinfo {author} {\bibfnamefont {Z.~K.}\ \bibnamefont
  {Liu}}, \bibinfo {author} {\bibfnamefont {J.}~\bibnamefont {Jiang}}, \bibinfo
  {author} {\bibfnamefont {B.}~\bibnamefont {Zhou}}, \bibinfo {author}
  {\bibfnamefont {Z.~J.}\ \bibnamefont {Wang}}, \bibinfo {author}
  {\bibfnamefont {Y.}~\bibnamefont {Zhang}}, \bibinfo {author} {\bibfnamefont
  {H.~M.}\ \bibnamefont {Weng}}, \bibinfo {author} {\bibfnamefont
  {D.}~\bibnamefont {Prabhakaran}}, \bibinfo {author} {\bibfnamefont {S.-K.}\
  \bibnamefont {Mo}}, \bibinfo {author} {\bibfnamefont {H.}~\bibnamefont
  {Peng}}, \bibinfo {author} {\bibfnamefont {P.}~\bibnamefont {Dudin}},
  \bibinfo {author} {\bibfnamefont {T.}~\bibnamefont {Kim}}, \bibinfo {author}
  {\bibfnamefont {M.}~\bibnamefont {Hoesch}}, \bibinfo {author} {\bibfnamefont
  {Z.}~\bibnamefont {Fang}}, \bibinfo {author} {\bibfnamefont {X.}~\bibnamefont
  {Dai}}, \bibinfo {author} {\bibfnamefont {Z.~X.}\ \bibnamefont {Shen}},
  \bibinfo {author} {\bibfnamefont {D.~L.}\ \bibnamefont {Feng}}, \bibinfo
  {author} {\bibfnamefont {Z.}~\bibnamefont {Hussain}}, \ and\ \bibinfo
  {author} {\bibfnamefont {Y.~L.}\ \bibnamefont {Chen}},\ }\href
  {http://www.nature.com/nmat/journal/v13/n7/full/nmat3990.html} {\bibfield
  {journal} {\bibinfo  {journal} {Nature Mater.}\ }\textbf {\bibinfo {volume}
  {13}},\ \bibinfo {pages} {677} (\bibinfo {year}
  {2014}{\natexlab{b}})}\BibitemShut {NoStop}%
\bibitem [{\citenamefont {Wan}\ \emph {et~al.}(2011)\citenamefont {Wan},
  \citenamefont {Turner}, \citenamefont {Vishwanath},\ and\ \citenamefont
  {Savrasov}}]{wan_topological_2011}%
  \BibitemOpen
  \bibfield  {author} {\bibinfo {author} {\bibfnamefont {X.}~\bibnamefont
  {Wan}}, \bibinfo {author} {\bibfnamefont {A.~M.}\ \bibnamefont {Turner}},
  \bibinfo {author} {\bibfnamefont {A.}~\bibnamefont {Vishwanath}}, \ and\
  \bibinfo {author} {\bibfnamefont {S.~Y.}\ \bibnamefont {Savrasov}},\ }\href
  {http://link.aps.org/doi/10.1103/PhysRevB.83.205101} {\bibfield  {journal}
  {\bibinfo  {journal} {Phys. Rev. B}\ }\textbf {\bibinfo {volume} {83}},\
  \bibinfo {pages} {205101} (\bibinfo {year} {2011})}\BibitemShut {NoStop}%
\bibitem [{\citenamefont {Hosur}\ and\ \citenamefont
  {Qi}(2013)}]{hosur_recent_2013}%
  \BibitemOpen
  \bibfield  {author} {\bibinfo {author} {\bibfnamefont {P.}~\bibnamefont
  {Hosur}}\ and\ \bibinfo {author} {\bibfnamefont {X.}~\bibnamefont {Qi}},\
  }\href {http://www.sciencedirect.com/science/article/pii/S1631070513001710}
  {\bibfield  {journal} {\bibinfo  {journal} {C. R. Phys.}\ }\textbf {\bibinfo
  {volume} {14}},\ \bibinfo {pages} {857} (\bibinfo {year} {2013})}\BibitemShut
  {NoStop}%
\bibitem [{\citenamefont {Xu}\ \emph {et~al.}(2015)\citenamefont {Xu},
  \citenamefont {Belopolski}, \citenamefont {Alidoust}, \citenamefont
  {Neupane}, \citenamefont {Zhang}, \citenamefont {Sankar}, \citenamefont
  {Huang}, \citenamefont {Lee}, \citenamefont {Chang}, \citenamefont {Wang},
  \citenamefont {Bian}, \citenamefont {Zheng}, \citenamefont {Sanchez},
  \citenamefont {Chou}, \citenamefont {Lin}, \citenamefont {Jia},\ and\
  \citenamefont {Hasan}}]{xu_experimental_2015}%
  \BibitemOpen
  \bibfield  {author} {\bibinfo {author} {\bibfnamefont {S.-Y.}\ \bibnamefont
  {Xu}}, \bibinfo {author} {\bibfnamefont {I.}~\bibnamefont {Belopolski}},
  \bibinfo {author} {\bibfnamefont {N.}~\bibnamefont {Alidoust}}, \bibinfo
  {author} {\bibfnamefont {M.}~\bibnamefont {Neupane}}, \bibinfo {author}
  {\bibfnamefont {C.}~\bibnamefont {Zhang}}, \bibinfo {author} {\bibfnamefont
  {R.}~\bibnamefont {Sankar}}, \bibinfo {author} {\bibfnamefont {S.-M.}\
  \bibnamefont {Huang}}, \bibinfo {author} {\bibfnamefont {C.-C.}\ \bibnamefont
  {Lee}}, \bibinfo {author} {\bibfnamefont {G.}~\bibnamefont {Chang}}, \bibinfo
  {author} {\bibfnamefont {B.}~\bibnamefont {Wang}}, \bibinfo {author}
  {\bibfnamefont {G.}~\bibnamefont {Bian}}, \bibinfo {author} {\bibfnamefont
  {H.}~\bibnamefont {Zheng}}, \bibinfo {author} {\bibfnamefont {D.~S.}\
  \bibnamefont {Sanchez}}, \bibinfo {author} {\bibfnamefont {F.}~\bibnamefont
  {Chou}}, \bibinfo {author} {\bibfnamefont {H.}~\bibnamefont {Lin}}, \bibinfo
  {author} {\bibfnamefont {S.}~\bibnamefont {Jia}}, \bibinfo {author}
  {\bibfnamefont {M.~Z.}\ \bibnamefont {Hasan}},\ }{\bibfield
  {journal} {\bibinfo  {journal} {arxiv:1502.03807v1}\ } }\BibitemShut {NoStop}%
\bibitem [{\citenamefont {Lu}\ \emph {et~al.}(2015)\citenamefont {Lu},
  \citenamefont {Wang}, \citenamefont {Ye}, \citenamefont {Ran}, \citenamefont
  {Fu}, \citenamefont {Joannopoulos},\ and\ \citenamefont
  {Solja\v{c}i\'{c}}}]{lu_experimental_2015}%
  \BibitemOpen
  \bibfield  {author} {\bibinfo {author} {\bibfnamefont {L.}~\bibnamefont
  {Lu}}, \bibinfo {author} {\bibfnamefont {Z.}~\bibnamefont {Wang}}, \bibinfo
  {author} {\bibfnamefont {D.}~\bibnamefont {Ye}}, \bibinfo {author}
  {\bibfnamefont {L.}~\bibnamefont {Ran}}, \bibinfo {author} {\bibfnamefont
  {L.}~\bibnamefont {Fu}}, \bibinfo {author} {\bibfnamefont {J.~D.}\
  \bibnamefont {Joannopoulos}}, \ and\ \bibinfo {author} {\bibfnamefont
  {M.}~\bibnamefont {Solja\v{c}i\'{c}}},\ }{\bibfield  {journal}
  {\bibinfo  {journal} {arxiv:1502.03438v1}\ } (\bibinfo {year}
  {2015})}\BibitemShut {NoStop}%
\bibitem [{\citenamefont {Lv}\ \emph {et~al.}(2015)\citenamefont {Lv},
  \citenamefont {Weng}, \citenamefont {Fu}, \citenamefont {Wang}, \citenamefont
  {Miao}, \citenamefont {Ma}, \citenamefont {Richard}, \citenamefont {Huang},
  \citenamefont {Zhao}, \citenamefont {Chen}, \citenamefont {Fang},
  \citenamefont {Dai}, \citenamefont {Qian},\ and\ \citenamefont
  {Ding}}]{lv_discovery_2015}%
  \BibitemOpen
  \bibfield  {author} {\bibinfo {author} {\bibfnamefont {B.~Q.}\ \bibnamefont
  {Lv}}, \bibinfo {author} {\bibfnamefont {H.~M.}\ \bibnamefont {Weng}},
  \bibinfo {author} {\bibfnamefont {B.~B.}\ \bibnamefont {Fu}}, \bibinfo
  {author} {\bibfnamefont {X.~P.}\ \bibnamefont {Wang}}, \bibinfo {author}
  {\bibfnamefont {H.}~\bibnamefont {Miao}}, \bibinfo {author} {\bibfnamefont
  {J.}~\bibnamefont {Ma}}, \bibinfo {author} {\bibfnamefont {P.}~\bibnamefont
  {Richard}}, \bibinfo {author} {\bibfnamefont {X.~C.}\ \bibnamefont {Huang}},
  \bibinfo {author} {\bibfnamefont {L.~X.}\ \bibnamefont {Zhao}}, \bibinfo
  {author} {\bibfnamefont {G.~F.}\ \bibnamefont {Chen}}, \bibinfo {author}
  {\bibfnamefont {Z.}~\bibnamefont {Fang}}, \bibinfo {author} {\bibfnamefont
  {X.}~\bibnamefont {Dai}}, \bibinfo {author} {\bibfnamefont {T.}~\bibnamefont
  {Qian}}, \ and\ \bibinfo {author} {\bibfnamefont {H.}~\bibnamefont {Ding}},\
  }{\bibfield  {journal} {\bibinfo  {journal}
  {arxiv:1502.04684v1}\ } (\bibinfo {year} {2015})}\BibitemShut {NoStop}%
\bibitem [{\citenamefont {Huang}\ \emph {et~al.}(2015)\citenamefont {Huang},
  \citenamefont {Zhao}, \citenamefont {Long}, \citenamefont {Wang},
  \citenamefont {Chen}, \citenamefont {Yang}, \citenamefont {Liang},
  \citenamefont {Xue}, \citenamefont {Weng}, \citenamefont {Fang},
  \citenamefont {Dai},\ and\ \citenamefont {Chen}}]{huang_observation_2015}%
  \BibitemOpen
  \bibfield  {author} {\bibinfo {author} {\bibfnamefont {X.}~\bibnamefont
  {Huang}}, \bibinfo {author} {\bibfnamefont {L.}~\bibnamefont {Zhao}},
  \bibinfo {author} {\bibfnamefont {Y.}~\bibnamefont {Long}}, \bibinfo {author}
  {\bibfnamefont {P.}~\bibnamefont {Wang}}, \bibinfo {author} {\bibfnamefont
  {D.}~\bibnamefont {Chen}}, \bibinfo {author} {\bibfnamefont {Z.}~\bibnamefont
  {Yang}}, \bibinfo {author} {\bibfnamefont {H.}~\bibnamefont {Liang}},
  \bibinfo {author} {\bibfnamefont {M.}~\bibnamefont {Xue}}, \bibinfo {author}
  {\bibfnamefont {H.}~\bibnamefont {Weng}}, \bibinfo {author} {\bibfnamefont
  {Z.}~\bibnamefont {Fang}}, \bibinfo {author} {\bibfnamefont {X.}~\bibnamefont
  {Dai}}, \ and\ \bibinfo {author} {\bibfnamefont {G.}~\bibnamefont {Chen}},\
  }{\bibfield  {journal} {\bibinfo  {journal}
  {arxiv:1503.01304v1}\ } (\bibinfo {year} {2015})}\BibitemShut {NoStop}%
\bibitem [{\citenamefont {Zhang}\ \emph {et~al.}(2015)\citenamefont {Zhang},
  \citenamefont {Xu}, \citenamefont {Belopolski}, \citenamefont {Yuan},
  \citenamefont {Lin}, \citenamefont {Tong}, \citenamefont {Alidoust},
  \citenamefont {Lee}, \citenamefont {Huang}, \citenamefont {Lin},
  \citenamefont {Neupane}, \citenamefont {Sanchez}, \citenamefont {Zheng},
  \citenamefont {Bian}, \citenamefont {Wang}, \citenamefont {Zhang},
  \citenamefont {Neupert}, \citenamefont {Hasan},\ and\ \citenamefont
  {Jia}}]{zhang_observation_2015}%
  \BibitemOpen
  \bibfield  {author} {\bibinfo {author} {\bibfnamefont {C.}~\bibnamefont
  {Zhang}}, \bibinfo {author} {\bibfnamefont {S.-Y.}\ \bibnamefont {Xu}},
  \bibinfo {author} {\bibfnamefont {I.}~\bibnamefont {Belopolski}}, \bibinfo
  {author} {\bibfnamefont {Z.}~\bibnamefont {Yuan}}, \bibinfo {author}
  {\bibfnamefont {Z.}~\bibnamefont {Lin}}, \bibinfo {author} {\bibfnamefont
  {B.}~\bibnamefont {Tong}}, \bibinfo {author} {\bibfnamefont {N.}~\bibnamefont
  {Alidoust}}, \bibinfo {author} {\bibfnamefont {C.-C.}\ \bibnamefont {Lee}},
  \bibinfo {author} {\bibfnamefont {S.-M.}\ \bibnamefont {Huang}}, \bibinfo
  {author} {\bibfnamefont {H.}~\bibnamefont {Lin}}, \bibinfo {author}
  {\bibfnamefont {M.}~\bibnamefont {Neupane}}, \bibinfo {author} {\bibfnamefont
  {D.~S.}\ \bibnamefont {Sanchez}}, \bibinfo {author} {\bibfnamefont
  {H.}~\bibnamefont {Zheng}}, \bibinfo {author} {\bibfnamefont
  {G.}~\bibnamefont {Bian}}, \bibinfo {author} {\bibfnamefont {J.}~\bibnamefont
  {Wang}}, \bibinfo {author} {\bibfnamefont {C.}~\bibnamefont {Zhang}},
  \bibinfo {author} {\bibfnamefont {T.}~\bibnamefont {Neupert}}, \bibinfo
  {author} {\bibfnamefont {M.~Z.}\ \bibnamefont {Hasan}}, \ and\ \bibinfo
  {author} {\bibfnamefont {S.}~\bibnamefont {Jia}},\ }{\bibfield
  {journal} {\bibinfo  {journal} {arxiv:1503.02630v1}\ } (\bibinfo {year}
  {2015})}\BibitemShut {NoStop}%
\bibitem [{\citenamefont {Burkov}\ and\ \citenamefont
  {Balents}(2011)}]{burkov_weyl_2011}%
  \BibitemOpen
  \bibfield  {author} {\bibinfo {author} {\bibfnamefont {A.~A.}\ \bibnamefont
  {Burkov}}\ and\ \bibinfo {author} {\bibfnamefont {L.}~\bibnamefont
  {Balents}},\ }\href {http://link.aps.org/doi/10.1103/PhysRevLett.107.127205}
  {\bibfield  {journal} {\bibinfo  {journal} {Phys. Rev. Lett.}\ }\textbf
  {\bibinfo {volume} {107}},\ \bibinfo {pages} {127205} (\bibinfo {year}
  {2011})}\BibitemShut {NoStop}%
\bibitem [{\citenamefont {Zhang}\ \emph {et~al.}(2005)\citenamefont {Zhang},
  \citenamefont {Tan}, \citenamefont {Stormer},\ and\ \citenamefont
  {Kim}}]{zhang_experimental_2005}%
  \BibitemOpen
  \bibfield  {author} {\bibinfo {author} {\bibfnamefont {Y.}~\bibnamefont
  {Zhang}}, \bibinfo {author} {\bibfnamefont {Y.-W.}\ \bibnamefont {Tan}},
  \bibinfo {author} {\bibfnamefont {H.~L.}\ \bibnamefont {Stormer}}, \ and\
  \bibinfo {author} {\bibfnamefont {P.}~\bibnamefont {Kim}},\ }\href
  {http://www.nature.com/nature/journal/v438/n7065/full/nature04235.html}
  {\bibfield  {journal} {\bibinfo  {journal} {Nature}\ }\textbf {\bibinfo
  {volume} {438}},\ \bibinfo {pages} {201} (\bibinfo {year}
  {2005})}\BibitemShut {NoStop}%
\bibitem [{\citenamefont {Tworzyd{\l}o}\ \emph {et~al.}(2006)\citenamefont
  {Tworzyd{\l}o}, \citenamefont {Trauzettel}, \citenamefont {Titov},
  \citenamefont {Rycerz},\ and\ \citenamefont
  {Beenakker}}]{tworzydlo_sub-poissonian_2006}%
  \BibitemOpen
  \bibfield  {author} {\bibinfo {author} {\bibfnamefont {J.}~\bibnamefont
  {Tworzyd{\l}o}}, \bibinfo {author} {\bibfnamefont {B.}~\bibnamefont
  {Trauzettel}}, \bibinfo {author} {\bibfnamefont {M.}~\bibnamefont {Titov}},
  \bibinfo {author} {\bibfnamefont {A.}~\bibnamefont {Rycerz}}, \ and\ \bibinfo
  {author} {\bibfnamefont {C.~W.~J.}\ \bibnamefont {Beenakker}},\ }\href
  {http://link.aps.org/doi/10.1103/PhysRevLett.96.246802} {\bibfield  {journal}
  {\bibinfo  {journal} {Phys. Rev. Lett.}\ }\textbf {\bibinfo {volume} {96}},\
  \bibinfo {pages} {246802} (\bibinfo {year} {2006})}\BibitemShut {NoStop}%
\bibitem [{\citenamefont {Katsnelson}(2006)}]{katsnelson_zitterbewegung_2006}%
  \BibitemOpen
  \bibfield  {author} {\bibinfo {author} {\bibfnamefont {M.~I.}\ \bibnamefont
  {Katsnelson}},\ }\href
  {http://link.springer.com/article/10.1140/epjb/e2006-00203-1} {\bibfield
  {journal} {\bibinfo  {journal} {Eur. Phys. J. B}\ }\textbf {\bibinfo {volume}
  {51}},\ \bibinfo {pages} {157} (\bibinfo {year} {2006})}\BibitemShut
  {NoStop}%
\bibitem [{\citenamefont {Baireuther}\ \emph {et~al.}(2014)\citenamefont
  {Baireuther}, \citenamefont {Edge}, \citenamefont {Fulga}, \citenamefont
  {Beenakker},\ and\ \citenamefont {Tworzyd{\l}o}}]{baireuther_quantum_2014}%
  \BibitemOpen
  \bibfield  {author} {\bibinfo {author} {\bibfnamefont {P.}~\bibnamefont
  {Baireuther}}, \bibinfo {author} {\bibfnamefont {J.~M.}\ \bibnamefont
  {Edge}}, \bibinfo {author} {\bibfnamefont {I.~C.}\ \bibnamefont {Fulga}},
  \bibinfo {author} {\bibfnamefont {C.~W.~J.}\ \bibnamefont {Beenakker}}, \
  and\ \bibinfo {author} {\bibfnamefont {J.}~\bibnamefont {Tworzyd{\l}o}},\
  }\href {http://link.aps.org/doi/10.1103/PhysRevB.89.035410} {\bibfield
  {journal} {\bibinfo  {journal} {Phys. Rev. B}\ }\textbf {\bibinfo {volume}
  {89}},\ \bibinfo {pages} {035410} (\bibinfo {year} {2014})}\BibitemShut
  {NoStop}%
\bibitem [{\citenamefont {Sbierski}\ \emph {et~al.}(2014)\citenamefont
  {Sbierski}, \citenamefont {Pohl}, \citenamefont {Bergholtz},\ and\
  \citenamefont {Brouwer}}]{sbierski_quantum_2014}%
  \BibitemOpen
  \bibfield  {author} {\bibinfo {author} {\bibfnamefont {B.}~\bibnamefont
  {Sbierski}}, \bibinfo {author} {\bibfnamefont {G.}~\bibnamefont {Pohl}},
  \bibinfo {author} {\bibfnamefont {E.~J.}\ \bibnamefont {Bergholtz}}, \ and\
  \bibinfo {author} {\bibfnamefont {P.~W.}\ \bibnamefont {Brouwer}},\ }\href
  {http://link.aps.org/doi/10.1103/PhysRevLett.113.026602} {\bibfield
  {journal} {\bibinfo  {journal} {Phys. Rev. Lett.}\ }\textbf {\bibinfo
  {volume} {113}},\ \bibinfo {pages} {026602} (\bibinfo {year}
  {2014})}\BibitemShut {NoStop}%
\bibitem [{\citenamefont {Beenakker}\ and\ \citenamefont
  {B\"uttiker}(1992)}]{beenakker_suppression_1992}%
  \BibitemOpen
  \bibfield  {author} {\bibinfo {author} {\bibfnamefont {C.~W.~J.}~\bibnamefont
  {Beenakker}}\ and\ \bibinfo {author} {\bibfnamefont {M.}~\bibnamefont
  {B\"uttiker}},\ }\href {http://link.aps.org/doi/10.1103/PhysRevB.46.1889}
  {\bibfield  {journal} {\bibinfo  {journal} {Phys. Rev. B}\ }\textbf {\bibinfo
  {volume} {46}},\ \bibinfo {pages} {1889} (\bibinfo {year}
  {1992})}\BibitemShut {NoStop}%
\bibitem [{\citenamefont {Jalabert}\ \emph {et~al.}(1994)\citenamefont
  {Jalabert}, \citenamefont {Pichard},\ and\ \citenamefont
  {Beenakker}}]{jalabert_universal_1994}%
  \BibitemOpen
  \bibfield  {author} {\bibinfo {author} {\bibfnamefont {R.~A.}\ \bibnamefont
  {Jalabert}}, \bibinfo {author} {\bibfnamefont {J.-L.}\ \bibnamefont
  {Pichard}}, \ and\ \bibinfo {author} {\bibfnamefont {C.~W.~J.}\ \bibnamefont
  {Beenakker}},\ }\href {http://iopscience.iop.org/0295-5075/27/4/001}
  {\bibfield  {journal} {\bibinfo  {journal} {Europhys. Lett.}\ }\textbf
  {\bibinfo {volume} {27}},\ \bibinfo {pages} {255} (\bibinfo {year}
  {1994})}\BibitemShut {NoStop}%
\bibitem [{\citenamefont {Baranger}\ and\ \citenamefont
  {Mello}(1994)}]{baranger_mesoscopic_1994}%
  \BibitemOpen
  \bibfield  {author} {\bibinfo {author} {\bibfnamefont {H.~U.}~\bibnamefont
  {Baranger}}\ and\ \bibinfo {author} {\bibfnamefont {P.~A.}~\bibnamefont
  {Mello}},\ }\href {http://link.aps.org/doi/10.1103/PhysRevLett.73.142}
  {\bibfield  {journal} {\bibinfo  {journal} {Phys. Rev. Lett.}\ }\textbf
  {\bibinfo {volume} {73}},\ \bibinfo {pages} {142} (\bibinfo {year}
  {1994})}\BibitemShut {NoStop}%
\bibitem [{\citenamefont {B\"uttiker}(1990)}]{buettiker_scattering_1990}%
  \BibitemOpen
  \bibfield  {author} {\bibinfo {author} {\bibfnamefont {M.}~\bibnamefont
  {B\"uttiker}},\ }\href {\doibase 10.1103/PhysRevLett.65.2901} {\bibfield
  {journal} {\bibinfo  {journal} {Phys. Rev. Lett.}\ }\textbf {\bibinfo
  {volume} {65}},\ \bibinfo {pages} {2901} (\bibinfo {year}
  {1990})}\BibitemShut {NoStop}%
\bibitem [{\citenamefont {Charlier}\ \emph {et~al.}(1991)\citenamefont
  {Charlier}, \citenamefont {Michenaud}, \citenamefont {Gonze},\ and\
  \citenamefont {Vigneron}}]{charlier_tight-binding_1991}%
  \BibitemOpen
  \bibfield  {author} {\bibinfo {author} {\bibfnamefont {J.-C.}\ \bibnamefont
  {Charlier}}, \bibinfo {author} {\bibfnamefont {J.-P.}\ \bibnamefont
  {Michenaud}}, \bibinfo {author} {\bibfnamefont {X.}~\bibnamefont {Gonze}}, \
  and\ \bibinfo {author} {\bibfnamefont {J.-P.}\ \bibnamefont {Vigneron}},\
  }\href {http://link.aps.org/doi/10.1103/PhysRevB.44.13237} {\bibfield
  {journal} {\bibinfo  {journal} {Phys. Rev. B}\ }\textbf {\bibinfo {volume}
  {44}},\ \bibinfo {pages} {13237} (\bibinfo {year} {1991})}\BibitemShut
  {NoStop}%
\bibitem [{\citenamefont {Pellegrino}\ \emph {et~al.}(2011)\citenamefont
  {Pellegrino}, \citenamefont {Angilella},\ and\ \citenamefont
  {Pucci}}]{pellegrino_transport_2011}%
  \BibitemOpen
  \bibfield  {author} {\bibinfo {author} {\bibfnamefont {F.~M.~D.}\
  \bibnamefont {Pellegrino}}, \bibinfo {author} {\bibfnamefont {G.~G.~N.}\
  \bibnamefont {Angilella}}, \ and\ \bibinfo {author} {\bibfnamefont
  {R.}~\bibnamefont {Pucci}},\ }\href
  {http://link.aps.org/doi/10.1103/PhysRevB.84.195404} {\bibfield  {journal}
  {\bibinfo  {journal} {Phys. Rev. B}\ }\textbf {\bibinfo {volume} {84}},\
  \bibinfo {pages} {195404} (\bibinfo {year} {2011})}\BibitemShut {NoStop}%
\bibitem [{\citenamefont {Trescher}\ and\ \citenamefont
  {Bergholtz}(2012)}]{trescher_flat_2012}%
  \BibitemOpen
  \bibfield  {author} {\bibinfo {author} {\bibfnamefont {M.}~\bibnamefont
  {Trescher}}\ and\ \bibinfo {author} {\bibfnamefont {E.~J.}\ \bibnamefont
  {Bergholtz}},\ }\href {http://link.aps.org/doi/10.1103/PhysRevB.86.241111}
  {\bibfield  {journal} {\bibinfo  {journal} {Phys. Rev. B}\ }\textbf {\bibinfo
  {volume} {86}},\ \bibinfo {pages} {241111} (\bibinfo {year}
  {2012})}\BibitemShut {NoStop}%
\bibitem [{\citenamefont {Bergholtz}\ \emph {et~al.}(2015)\citenamefont
  {Bergholtz}, \citenamefont {Liu}, \citenamefont {Trescher}, \citenamefont
  {Moessner},\ and\ \citenamefont {Udagawa}}]{bergholtz_topology_2015}%
  \BibitemOpen
  \bibfield  {author} {\bibinfo {author} {\bibfnamefont {E.~J.}\ \bibnamefont
  {Bergholtz}}, \bibinfo {author} {\bibfnamefont {Z.}~\bibnamefont {Liu}},
  \bibinfo {author} {\bibfnamefont {M.}~\bibnamefont {Trescher}}, \bibinfo
  {author} {\bibfnamefont {R.}~\bibnamefont {Moessner}}, \ and\ \bibinfo
  {author} {\bibfnamefont {M.}~\bibnamefont {Udagawa}},\ }\href
  {http://link.aps.org/doi/10.1103/PhysRevLett.114.016806} {\bibfield
  {journal} {\bibinfo  {journal} {Phys. Rev. Lett.}\ }\textbf {\bibinfo
  {volume} {114}},\ \bibinfo {pages} {016806} (\bibinfo {year}
  {2015})}\BibitemShut {NoStop}%
\bibitem [{\citenamefont {Katayama}\ \emph {et~al.}(2006)\citenamefont
  {Katayama}, \citenamefont {Kobayashi},\ and\ \citenamefont
  {Suzumura}}]{katayama_pressure-induced_2006}%
  \BibitemOpen
  \bibfield  {author} {\bibinfo {author} {\bibfnamefont {S.}~\bibnamefont
  {Katayama}}, \bibinfo {author} {\bibfnamefont {A.}~\bibnamefont {Kobayashi}},
  \ and\ \bibinfo {author} {\bibfnamefont {Y.}~\bibnamefont {Suzumura}},\
  }\href {http://journals.jps.jp/doi/abs/10.1143/JPSJ.75.054705} {\bibfield
  {journal} {\bibinfo  {journal} {J. Phys. Soc. Jpn.}\ }\textbf {\bibinfo
  {volume} {75}},\ \bibinfo {pages} {054705} (\bibinfo {year}
  {2006})}\BibitemShut {NoStop}%
\end{thebibliography}
%

\end{document}